\documentstyle[12pt,aaspp4,astrobib,psfig]{article} 

\newcommand{\etal} {{\it et~al.\ }}

\begin{document}

\title{Drizzle:  A Method for the Linear Reconstruction of Undersampled Images} 


\author{A. S. Fruchter$^1$ and R. N. Hook$^2$} 

\affil{$^{1}$Space Telescope Science Institute, 3700 San Martin
Drive, Baltimore, MD 21218, USA; fruchter@stsci.edu\\
$^2$Space Telescope European Coordinating Facility, Karl Schwarzschild Strasse 2, D-85748 Garching, Germany; rhook@eso.org}


\begin{abstract}
We have developed a method for the linear reconstruction of an image
from undersampled, dithered data. 
The algorithm, known
as Variable-Pixel Linear Reconstruction, or informally as
``Drizzle'', preserves photometry and resolution, can weight input
images according to the statistical significance of each pixel, and
removes the effects of geometric distortion both on image shape and
photometry.  This paper presents the method and its implementation. 
The photometric and astrometric
accuracy and image
fidelity of the algorithm as well as the noise
characteristics of output images are discussed.  
In addition, we describe the
use of drizzling to combine dithered images in the presence of cosmic
rays.
\end{abstract}



\section{INTRODUCTION} 

Undersampled images are common in astronomy, because instrument
designers are frequently forced to choose between properly sampling
a small field-of-view, or undersampling a larger field.  Nowhere
is this problem more acute than on the Hubble Space Telescope,
whose corrected optics now
provide superb resolution;
however, the detectors
on HST are only able to take full advantage of the full resolving power
of the telescope over a limited field of view.  
For instance, the primary optical imaging camera on the HST, the Wide
Field and Planetary Camera~2 \cite{tbbc+94}, is composed of four separate
800x800 pixel CCD cameras, one of which, the planetary
camera (PC) has a scale of $0\farcs046$ per pixel, while the other three,
arranged in a chevron around the PC, have a scale of $0\farcs097$ per
pixel.  These latter three cameras, referred to as the wide field
cameras (WFs), are currently
the primary workhorses for deep imaging surveys on HST.
However, these cameras greatly undersample the HST image.  The width
of a WF pixel
equals the full-width at half-maximum (FWHM) of the optics in the the 
near-infrared, and greatly exceeds
it in the blue.   In contrast, a well-sampled detector would have
$\ga 2.5$ pixels across the FWHM.
Other HST cameras such as NICMOS, STIS and the future Advanced Camera
for Surveys (ACS) also suffer from undersampling to varying degrees.
The effect of undersampling
on WF images is illustrated by the ``Great Eye Chart in the
Sky" in Figure 1. Further examples showing astronomical targets are given
in Section 8. 

\begin{figure}
\centerline{\psfig{figure=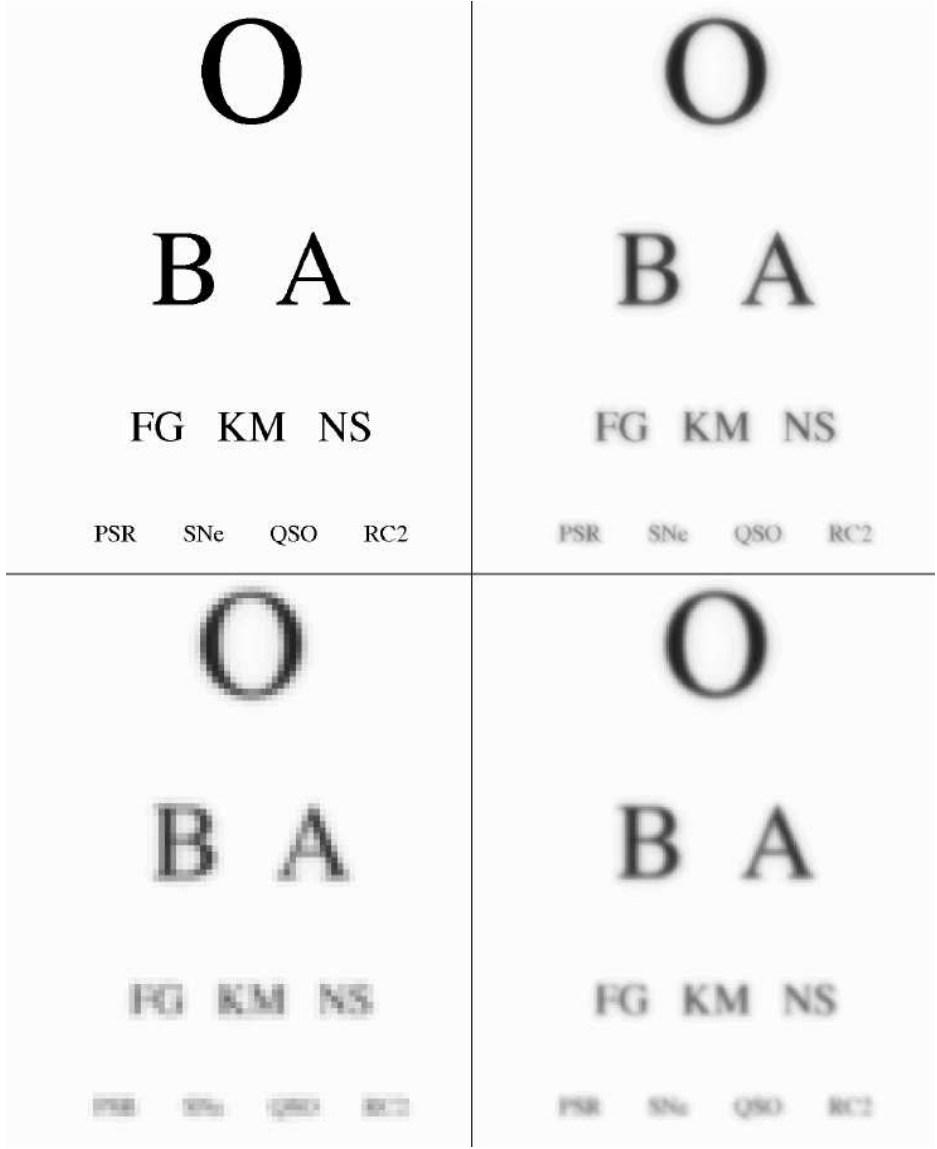,height=6.0in}}
\caption{In the upper left corner of this figure, we present 
the ``true image'',
{\it i.e.} the image one would see with an infinitely large telescope.  The
upper right shows the image after convolution with the optics of the
Hubble Space Telescope and the
WFPC2 camera -- the primary wide-field imaging instrument presently
installed on the HST.  The lower
left shows the image after sampling by the WFPC2 CCD, and the lower right
shows a linear reconstruction of dithered CCD images.}
\end{figure}

When the true distribution of light on the sky $T$ is observed by a telescope
it is  convolved by the point-spread function of the optics $O$ to 
produce an observed image, 
$ I_o = T \otimes O $, where $\otimes$ represents the convolution operator.  
This effect is shown for the HST
and WFPC2 optics by the upper-right panel in
Figure~1.  Pixelated detectors then again convolve this image with the
response function of the electronic pixel $E$, thus
$I_d = T \otimes O \otimes E$.  The detected image can be thought
of as this continuous convolved image {\it sampled} at the center
of each physical pixel.  Thus a shift in the position of the detector
(know as a ``dither'') can be thought of
as producing offset samples from the same convolved image.
Although pixels are typically
square on the detector, their response may be non-uniform, and indeed,
may, because of the scattering of light and charge carriers,   effectively 
extend beyond the physical pixel boundaries.  This is the case in WFPC2.
By contrast, in the NICMOS detectors \cite{shmh+99,lau99}, the electronic pixel
is effectively smaller than the physical pixel.

Fortunately, much of the information lost to undersampling
can be restored.  In the lower right of Figure 1
we display an image made using one of the family of techniques we refer
to as ``linear reconstruction.''  The most commonly used of these 
techniques are
shift-and-add and interlacing.
In interlacing, the pixels from the independent images are 
placed in alternating pixels on the output image according to
the alignment of the pixel centers in the original images.
The image in the lower right corner of Figure 1 has been restored by
interlacing dithered images.  
However, due to the occasional small positioning errors of 
the telescope and the 
non-uniform shifts in pixel space caused by the 
geometric distortion of the optics, true interlacing of
 images is
often infeasible.  In the other standard linear reconstruction technique,
shift-and-add, a pixel is shifted to the appropriate location 
and then added onto a sub-sampled image.

Shift-and-add
 can easily handle arbitrary dither postions, but it
convolves the image yet again with the original pixel, compounding the
blurring of the
image and the correlation of the noise.
In this case, two further
convolutions are involved.  
The image is convolved with the physical
pixel $P$, as this pixel is mathematically shifted over and added
to the final image.  In addition, when many images with different pointings are
added together on the final output grid, there is also a 
convolution by the pixel of the final output grid $G$.   This produces
a final image
\begin{equation}
I = T \otimes O \otimes E \otimes P \otimes G .
\end{equation} 
The last convolution rarely produces a significant effect, however, as
the output grid is usually considerably finer than the detector pixel
grid, and convolutions add roughly as a sum of squares (the summation
is exact if the convolving functions are Gaussians).

The importance of avoiding convolutions
by the detector pixel is emphasized by comparing the upper and lower
right
hand images in Figure 1.
The deterioration in image quality between these images
is due entirely to the single convolution of the image by the WF pixel.
The interlaced image in the lower-right panel has had the sampled 
values from all of the input images directly placed in the appropriate
output pixels without further convolution by either $P$ or $G$.

Here we present a new method, Drizzle, which was originally
developed for combining the dithered images of the Hubble 
Deep Field North (HDF-N) \cite{hdf+96} and has since been widely used both
for the combination of dithered images from HST's cameras and those
on other telescopes.
Drizzle has the versatility of shift-and-add yet largely maintains
the resolution and independent noise statistics of interlacing.
While other methods ({\it c.f.} Lauer 1999) have \nocite{lau99} 
been suggested for 
the linear combination of dithered images, Drizzle has the advantage
of being able to handle images with essentially arbitrary shifts, 
rotations and geometric
distortion, and, when given input images with proper associated weight maps,
creates an optimal statistically summed image.   Drizzle also naturally
handles images with ``missing'' data, due, for instance, to corruption
by cosmic rays or detector defects.

The reader should note that Drizzle does not
attempt to improve upon the final image
resolution by enhancing the high frequency components of the
image which have been suppressed either by the optics or the
detector.  While such procedures, which we refer to as ``image
restoration'' (in contrast to ``image reconstruction''),
are frequently very valuable (see Hanisch and White 1994 \nocite{hw94}
for a review), they
invariably trade signal-to-noise for enhanced resolution.   Drizzle,
on the other hand,
was developed specifically to provide a flexible and general method of
image combination which produces
high resolution results without sacrificing the final signal-to-noise.

  \section{THE METHOD}

Although the effect of Drizzle on the quality
of the image can be profound, the
algorithm is conceptually straightforward. Pixels
in the original input images
are mapped into pixels in the subsampled output image, taking into account
shifts and rotations between images and the optical distortion of the camera.
However, in order to avoid 
re-convolving the image with the large pixel ``footprint"
of the camera, we allow the user to shrink the pixel before it is averaged
into the output image, as shown in Figure 2.

\begin{figure}
\centerline{\psfig{figure=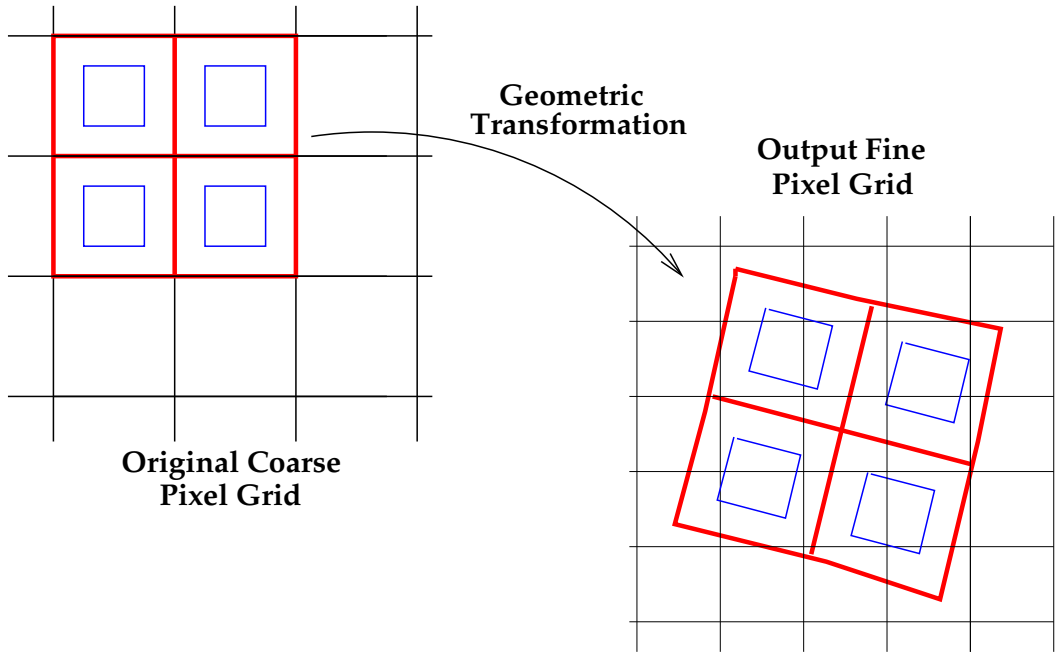,width=5.5in}}
\caption{A schematic representation of Drizzle.  The input pixel grid
(shown on the left) is mapped onto a finer output grid (shown on right),
taking into accounts shift, rotation and geometric distortion.  The user
is allowed to ``shrink" the input pixels to smaller pixels, which we refer to
as drops (faint inner squares).   A given input image only affects
output image pixels under drops.  In this particular case, the central
output pixel receives no information from the input image.}
\end{figure}

The new shrunken pixels, or ``drops",  rain
down upon the subsampled output.
In the case of the HDF-N WFPC2 imaging, the drops were given
linear dimensions one-half that of the input pixel --- slightly larger than
the dimensions of the output pixels. 
The value of an input pixel is averaged into an output pixel with 
a weight proportional to the area of overlap between the ``drop" and
the output pixel.
Note that if the drop size is sufficiently small
not all output pixels have data added to them from each
input image.   One must therefore choose a drop size that is small
enough to avoid degrading the image, but large enough so that 
after all images are drizzled the coverage is reasonably uniform.

The drop size is controlled by a user-adjustable parameter called
{\tt pixfrac}, which is simply the ratio of the linear size of the drop
to the input pixel (before any adjustment due to the geometric distortion
of the camera).   Thus interlacing is equivalent to Drizzle in the limit of 
${\rm{\tt pixfrac}} \rightarrow 0.0$,
while shift-and-add is equivalent to ${\rm{\tt pixfrac}}=1.0$.  The degree
of subsampling of the output is controlled by the user through the
scale parameter $s$, which is the ratio of the linear size of an output
pixel to an input pixel.

When a pixel $(x_i,y_i)$ from an input image $i$ with data value
$d_{x_i y_i}$ and user defined weight $w_{x_iy_i}$ is added 
to an output image pixel $(x_o,y_o)$
with value $I_{x_o y_o}$, weight $W_{x_o y_o}$,
and fractional pixel overlap $0 < a_{x_i y_i x_o y_o} < 1$, the resulting
values and weights of that same pixel, $I'$ and $W'$ are
\begin{eqnarray}
W'_{x_o y_o} &=& a_{x_i y_i x_o y_o}w_{x_i y_i} + W_{x_o y_o}\\
I'_{x_o y_o} &=& \frac{d_{x_i y_i}a_{x_i y_i x_o y_o}w_{x_i y_i}s^2 + I_{x_o y_o}W_{x_o y_o}}{W'_{x_o y_o}}
\end{eqnarray}
where a factor of $s^2$ is introduced to conserve surface intensity,
and where $i$ and $o$ are used to distinguish the input and output
pixel indices.
In practice, Drizzle applies this iterative procedure to the
input data, pixel-by-pixel, image-by-image.  Thus, after
each input image is processed, there is a usable output
image and weight, $I$ and $W$. 

The final output images, after all inputs have been processed, can be
written as 
\begin{eqnarray}
W_{x_o y_o} &=& a_{x_i y_i x_o y_o}w_{x_i y_i} \\
I_{x_o y_o} &=& \frac{d_{x_i y_i}a_{x_i y_i x_o y_o}w_{x_i y_i}s^2}{W_{x_o y_o}}
\end{eqnarray}
where for these Equations, 4 and 5, we use the Einstein
convention of summation over 
repeated indices, and where the input indices $x_i$ and $y_i$ extend over
all input images.  It is worth noting that
in nearly all cases $a_{x_i y_i x_o y_o} = 0$,
since very few input pixels overlap a given output pixel.

This algorithm has a number of advantages over the more standard
linear reconstruction methods presently used.
It preserves both absolute surface 
and point source photometry (though see Section 5 for a more detailed
discussion of point source photometry).   Therefore
flux can be measured using an aperture whose size is
independent of position on the chip.  
And as the method anticipates that  a given output pixel may receive
no information from a given input pixel, missing data (due for
instance to cosmic rays or detector defects) does not cause
a substantial problem, so long as there are enough dithered
images to fill in the gaps caused by these zero-weight input pixels.
Finally, the linear weighting scheme is
statistically optimum when inverse variance maps are used as the
input weights.

Drizzle replaces the convolution by $P$ in Equation~1 with a
convolution with $p$, the {\tt pixfrac}.  As this kernel is usually
smaller the full pixel, and as noted earlier convolutions add
as the sum of squares, the effect of this replacement is often
quite significant.  Furthermore, when the dithered positions of
the input images map directly onto the centers of the output
grid, and {\tt pixfrac} and {\tt scale} are chosen so that $p$ is 
only slightly greater than  $s$,
one obtains the full advantages of interlacing:  
because 
the power in an output pixel is almost entirely determined by input
pixels centered on that output pixel,
the convolutions with both $p$ and $G$ effectively drop away.
Nonetheless, the small overlap between adjacent drops
fills in missing data. 


\section{COSMIC RAY DETECTION}
 
Few HST observing proposals have sufficient time to take a number of
exposures at each of several dither positions.
Therefore, if dithering is to be of wide-spread use, one must be
able to remove cosmic rays from data where few, if any,
images are taken at the same position on the sky. We have therefore
adapted Drizzle to the removal of cosmic rays.  
As the techniques involved in cosmic ray removal are also valuable
in characterizing the image fidelity of Drizzle, we will
discuss them first.

Here then  is short description of the method we use for the
removal of cosmic rays:

\begin{enumerate}
\item{Drizzle each image onto a separate sub-sampled output image using 
${\rm{\tt pixfrac}}=1.0$}.
 
\item{Take the median of the resulting aligned drizzled images.   
This provides a first estimate of an image free of cosmic-rays.}
 
\item{Map the median image back to the input plane of each of the 
individual images, taking into account the
image shifts and geometric distortion.  This can done by interpolating
the values of the median image using a program
we have named ``Blot''.}  
 
\item{Take the spatial derivative of each of the blotted output images.
This derivative image is used in the next step to estimate the degree to which
errors in the computed image shift or the blurring effect of taking
the median could have distorted the value of the blotted estimate.}
 
\item{Compare each original image with the corresponding blotted image.  
Where the difference
is larger than can be explained by noise statistics, the flattening effect 
of taking the median or an error in the shift,
the suspect pixel is masked.}
 
\item{Repeat the 
previous step on pixels adjacent to pixels already masked, using
 a more stringent comparison criterion.}

\item{Finally, drizzle the input images onto a single output image
using the pixel masks created in the previous steps.
For this final combination a smaller {\tt pixfrac}
than in step 1) will usually be used
in order to maximize the resolution of the final image.
}

\end{enumerate} 

Figure 3 shows the result of applying this method to data originally
taken by Cowie and colleagues \cite{chs95}.   The reduction was done
using a set of IRAF \cite{tody93} scripts which
are now available along with Drizzle in the Dither package of STSDAS.  
In addition to demonstrating
how effectively cosmic rays can be removed from singly dithered images
(i.e. images which share no common pointing), this image also
displays the degree to which linear reconstruction can improve the detail
of an image.  In the Drizzled image the object to the upper right clearly
has a double nucleus (or a single nucleus with a dust lane through it), but
in the original image the object appears unresolved.  

\begin{figure}[t]

\centerline{\psfig{figure=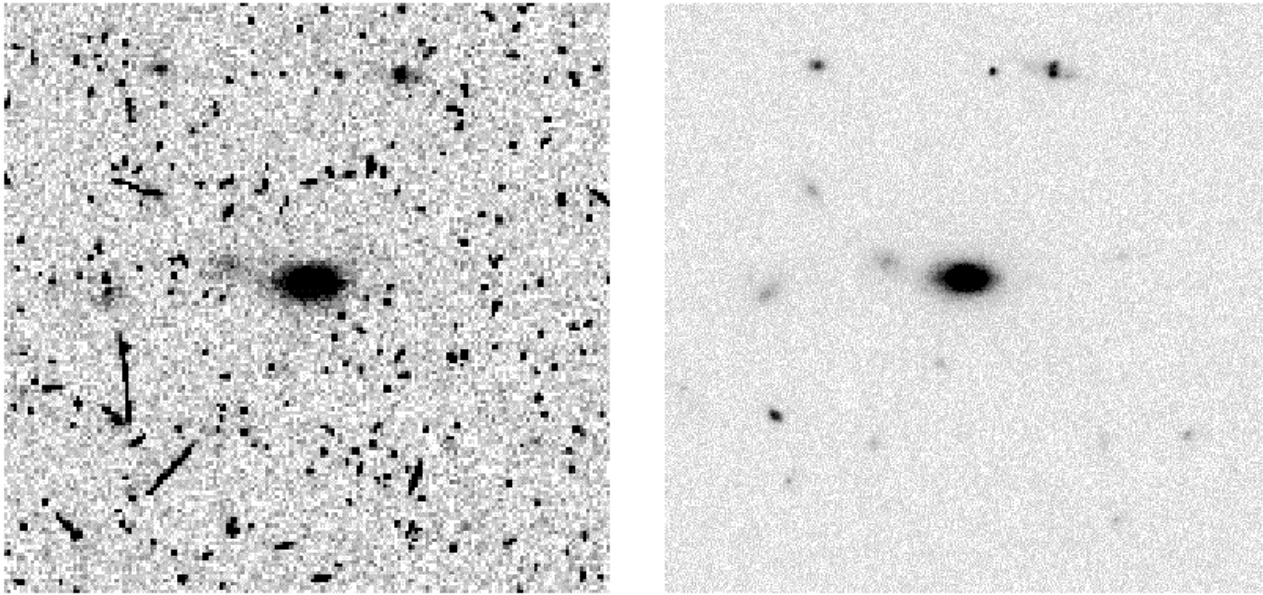,width=6.75in}}

\caption{The image on the left shows a region
of one of twelve 2400s archival images taken
with the F814W wide near-infrared filter on WFPC2.
Numerous cosmic
rays are visible.  On the right is the drizzled combination of the twelve
images, no two of which shared a dither position.}
\end{figure}

\section{IMAGE FIDELITY}
The drizzling algorithm was designed to obtain optimal signal-to-noise
on faint objects while preserving image resolution.
These goals are unfortunately not fully compatible.  
As noted earlier, non-linear
image restoration procedures which attempt to remove the blurring due to 
the PSF and the pixel response function
through enhancing the high frequencies in the image, such
as the Richardson-Lucy \cite{ric72,luc74,lh91}
and maximum entropy methods \cite{gd78,wd90}, directly exchange signal-to-noise 
for resolution.
In the drizzling algorithm no compromises on signal-to-noise have been made; 
the weight of an input pixel in the final output image is entirely
independent of its position on the chip.   Therefore, if the dithered images
do not uniformly sample the
field, the ``center of light'' in an output pixel may be offset from the center
of the pixel, and that offset may vary between adjacent pixels.  
Dithering offsets 
combined
with geometric distortion generally
produce a sampling pattern that varies across the
field.
The output PSFs produced by the combination of such irregularly
dithered datasets may, on occasion, show 
variations about the true PSF. 
Fortunately this
does not noticeably affect aperture photometry performed with typical
aperture sizes.    In practice the variability 
appears larger in WFPC2 data
than we would predict based on our simulations. 
Examination of more recent dithered stellar fields
leads us to suspect that this excess variability results 
from a problem with the
original data, possibly caused by charge transfer errors in the CCD 
\cite{wh97}.

\section{PHOTOMETRY}
 
Camera optics generally introduce geometric distortion of images.
In the case of the WFPC2, pixels at the corner
of each CCD subtend less area on the sky than those near the center.
This effect will be even more pronounced in the case of the Advanced
Camera for Surveys (ACS). 
However, after application of the flat field, a source of uniform surface
brightness on the sky  produces uniform counts across the CCD.  Therefore,
point sources near the corners of the chip are artificially brightened
compared to those in the center.  By scaling the weights of the
input pixels by their areal overlap with the output pixel, and by moving
input points to their corrected geometric positions, Drizzle largely
removes this effect.  In the case of ${\tt pixfrac} = 1$,
this correction is exact.

In order to study the ability of Drizzle to remove the photometric 
effects of geometric distortion when {\tt pixfrac} is not
identically equal to one,
we created a 
four times sub-sampled grid of
$19 \times 19$ artificial stellar PSFs.  This image was was then blotted
onto four separate images, each with the original WF sampling, but
dithered in a four-point pattern of half-pixel shifts.  As a result
of the geometric distortion of the WF camera, the stellar images in
the corners of these images appear up to $\sim 4 \%$ brighter in the
corners of the images than near the center.  These images were then
drizzled with a ${\tt scale}=0.5$ and  ${\tt pixfrac}=0.6$.  
Aperture photometry on the $19 \times 19$ grid after drizzling reveals that
the effect of geometric distortion on the photometry has been dramatically
reduced: the RMS photometric variation in the drizzled image is 0.004 mags.
Of course this is not the final photometric error of a drizzled image
(which will depend on the quality of the input images), but only
the additional error which the use of Drizzle would add under these
rather optimal circumstances.

\begin{figure}

\centerline{\psfig{figure=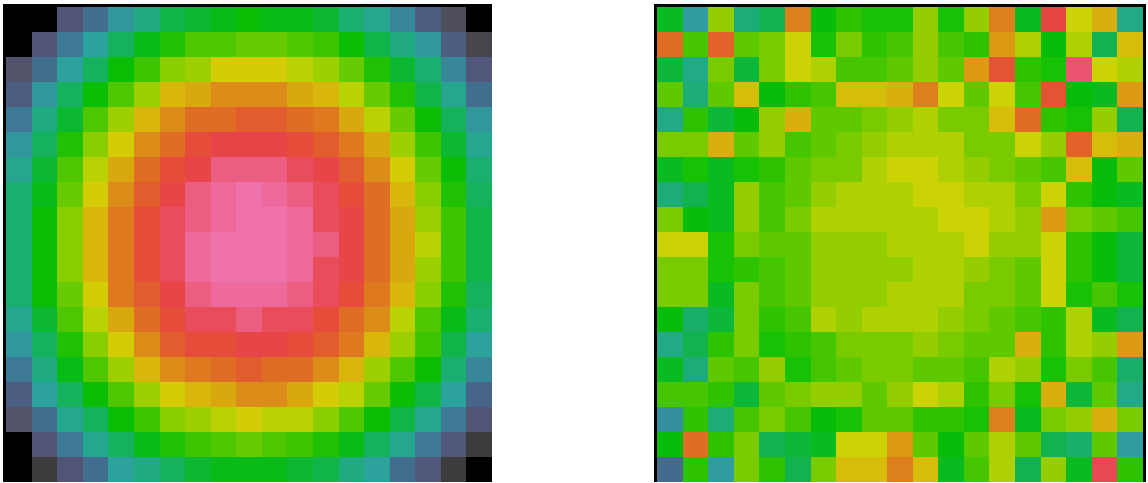,width=6.0in,angle=-90}}
\caption{An example of the
effect of geometric distortion on photometry.  The pixels
in the two images represent the photometric values of a $19 \times 19$ grid
of stars on the WF chip.  Each pixel represents the photometric value
of a star;  the pixels are not images of the stars.
The figure on the left shows the apparent brightness of the stars (all
of equal intrinsic brightness) as they would appear in a flat-fielded
WF image.
As pixels near the edge of the chip are up to 4\% smaller on the
sky than pixels near the center, the flat field has
artificially brightened the stars near the edges and the corners.
The image on the right shows the photometric values of these stars after
drizzling, including the use of representative cosmic ray masks (see
Figure 3).  The standard deviation of the corrected stellar magnitudes
$\le 0.015$~mags.}
\end{figure}

In practice users may not have four relatively well interlaced
images but rather a  number of almost random dithers, and
each dithered image may suffer from cosmic ray hits.
Therefore, in a separate simulation,
we have used the shifts actually obtained
in the WF2 F814W images of the HDF-N as an example of 
the nearly random sub-pixel phase that large dithers may produce
on HST.  In addition, we have associated with each image
a mask corresponding to cosmic ray hits from one of the deep HST
WF images used in creating Figure 3. 
When these simulated images
 are drizzled together, the root mean-square noise 
in the final photometry (which does not include any errors
that could occur because of missed or incorrectly identified cosmic
rays) is $\lesssim 0.015$ mags.   Figure~4 displays the results
of this process.

\section{ASTROMETRY}

We have also evaluated the effect of drizzling on astrometry.
The stellar images described in the previous section were again drizzled
using the HDF shifts 
as above, setting 
${\tt scale}=0.5$ and ${\tt pixfrac}=0.6$.  Both
uniform weight files and cosmic ray masks were used.  
The positions of the drizzled
stellar images were then determined with the ``imexam" task of 
IRAF,
which locates the centroid using the marginal statistics of a box about the
star.  A box with side equal to 6 {\em output} pixels, or slightly 
larger than 
twice the full-width at half maximum of the stellar images, was used.
A root mean square scatter of the stellar positions of $\sim 0.018$ 
{\em input}
pixels about the true position
was found for the images created with uniform weight files and 
the cosmic-ray masks.
However, we find an identical scatter when we down-sample the original
four-times oversampled images to the two-times oversampled scale of
the test images.   Thus it appears that {\it no} additional measurable
astrometric error has been introduced by Drizzle.  Rather we are simply
observing the limitations of our ability to centroid on images which
contain power that is not fully Nyquist sampled
even when using pixels half the original size.

\section{NOISE IN DRIZZLED IMAGES}

\subsection{The Nature of the Problem}

Drizzle frequently divides the power from a given input pixel between
several output pixels.  As a result, the noise in adjacent pixels will
be correlated.   
Understanding this effect in a quantitative manner is essential for
estimating the statistical errors when drizzled images are analysed using
object detection and measurement programs such as SExtractor \cite{sex-ref}
and DAOPHOT \cite{ste87}. 

The correlation of adjacent pixels implies that 
a measurement of the noise in a drizzled image
on the output pixel scale
underestimates the noise on larger scales. 
In particular, if one block sums a drizzled image by ${\rm N} \times {\rm
N}$ pixels, even using a proper weighted sum of the pixels,
the per-pixel noise
in the block summed image will
generally be more than a factor of N greater than the per-pixel noise of
the original image.  
The factor by which the ratio of these noise values 
differs
from N in the limit as ${\rm N} \Rightarrow \infty$ we refer to as the
noise correlation ratio, $\cal R$.

\begin{figure}[h]
\centerline{\psfig{figure=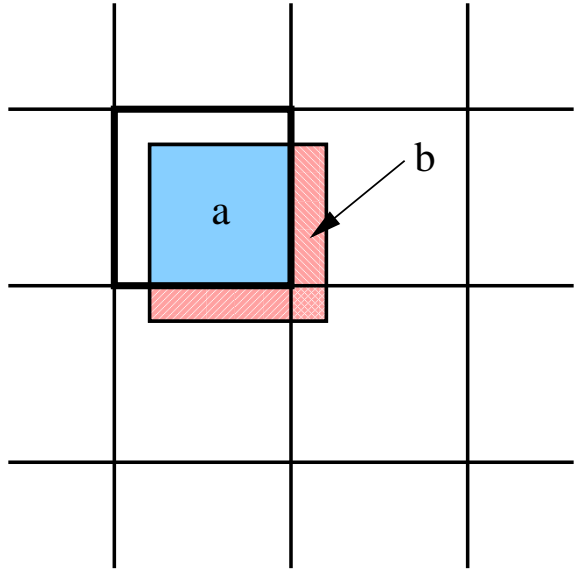,height=3.0in}}
\caption{ A schematic view of the distribution of noise from a single
input between neighboring output pixels.  A
more complete description is provided in the body of the paper.}
\end{figure}

One can easily see how this situation arises by examining
Figure 5.  In this figure we show an input pixel (broken up
into two regions, a and b) being drizzled onto an output
pixel plane.  Let the noise in this pixel be $\epsilon$ and let
the area of overlap of the drizzled pixel with the ``primary''
output pixel (shown with a heavier border) be $a$, and the areas
of overlap with the other three pixels be $b_1,b_2,$ and $b_3$, where
$b= b_1 + b_2 + b_3$, and $a + b = 1$.   Now the total noise power
added to the image variance is, of course, $\epsilon ^2$; however,
the noise that one would measure by simply adding up the variance
of the output image pixel-by-pixel  would be 
\begin{displaymath}
(a^2 + b_1^2 + b_2^2 + b_3^3) \epsilon^2 < \epsilon^2.
\end{displaymath}
The inequality exists because all cross terms ($ab_1, ab_2, b_1b_2 ....$)
are missed by summing the squares of the
individual pixels.  These terms, which represent
the correlated noise in a drizzled image,
can be significant.   

\subsection{The Calculation}

In general, the correlation between pixels, and thus $\cal R$, 
depends on the
choice of drizzle parameters and geometry and orientation of the
dither pattern, and often varies across an image.   While
it is always possible to estimate $\cal R$ for a given set of
drizzle parameters and dithers, in the case where all output
pixels receive equivalent inputs (in both dither pattern and noise,
though not necessarily from the same input images) the situation
becomes far more analytically tractable.  In this case, calculating
the noise properties of a single pixel gives one the noise properties
of the entire image.

Consider then the situation when ${\tt pixfrac}$, $p$, is set to zero.  
There is then no correlated noise in the output image 
-- since a given input pixel contributes only to the output
pixel which lies under its center, and the noise in the individual
input pixels is assumed to be independent.  
Let $d_{xy}$ represent a pixel from any of the input images,
and let
${\cal C}$ be the set of all
$d_{xy}$
whose centers fall on a given output pixel of interest. 
Then it is simple to
show from Equations 4 and 5 that the 
expected variance of the noise in that output pixel,
when $p=0$, is simply     
\begin{equation}
\sigma_c^2 = {\sum_{d_{xy} \in {\cal C}} w_{xy}^2 s^4 \sigma_{xy}^2} /{\left( \sum_{d_{xy} \in {\cal C}} w_{xy} \right)^2}
\end{equation}
where $\sigma_{xy}$ is the standard deviation of the noise distribution of the
input pixel $d_{xy}$.  We term this $\sigma_c$, as it is the standard deviation
calculated with the pixel values added only to the pixels on which they
are {\it centered}.

Now let us consider a drizzled output
image where $p > 0$.  In this case, the set of pixels contributing
to an output pixel will not only include pixels whose centers fall on 
the output pixel,
but also those for which a portion of the drop lands on the
output pixel of interest even though the center does not.    
We refer to the set of all input
pixels whose drops overlap with a given output pixel as ${\cal P}$
and note that ${\cal C} \subset {\cal P}$.  
The variance of the
noise in a given output
pixel is then
\begin{equation}
\sigma_p^2 = {\sum_{d_{xy} \in {\cal P}} a^2_{xy}w_{xy}^2 s^4 \sigma_{xy}^2} /{\left( \sum_{d_{xy} \in {\cal P}} a_{xy}w_{xy} \right)^2}
\end{equation}
where $a_{xy}$ is the fractional area overlap of the drop of input 
data pixel $d_{xy}$
with the output pixel $o$.   Here we choose the symbol $\sigma_p$ to represent
the standard deviation calculated from all pixels that contribute to
the output pixel when ${\tt pixfrac} = p$.
The degree to which $\sigma_p^2$ and $\sigma_c^2$ differ depends
on the dither pattern and the values of $p$ and $s$. 
However, as more input pixels are averaged together 
to estimate the value of a given output pixel in ${\cal P}$  than in
${\cal C}$, 
 $\sigma_p^2 \le \sigma_c^2$.  When
$p = 0$, $\sigma_p$ is by definition equal to $\sigma_c$. 

Now consider the situation where we block average a region of $\rm{N} \times
\rm{N}$ pixels of the final drizzled image, 
doing a proper weighted sum of the image pixels.  This
sum is equivalent to having drizzled onto an output image with
a scale size ${\rm N} s$.  But as ${\rm N}s \gg p$, this 
approaches  the sum over {\cal C}, or, in the limit of large
$N$, $N\sigma_c$.  However, a prediction of the noise in this
region, based solely on a measurement of the pixel-to-pixel noise, without
taking into account the correlation between pixels would produce
$N\sigma_p$.  Thus we see that 

\begin{displaymath}
{\cal R} = \frac{\sigma_c}{\sigma_p}
\end{displaymath}

One can therefore obtain $\cal R$ for a given set of drizzle parameters
and dither pattern by calculating $\sigma_c$ and $\sigma_p$ and performing
the division.  However, there is a further simplification that can be
made.  Because we have assumed that the inputs to each pixel are statistically
equivalent, it follows that the weights of the individual output
pixels in the
final drizzled image are independent of the choice of $p$.  
To see this, notice
that the total weight of a final image (that is the sum of the weights
of all of the pixels in the final image) is independent of the choice of
$p$.  Ignoring edge pixels, the number of pixels in the final image
with non-zero weight is also
independent of the choice of $p$.  Yet as the fraction
of pixels within $p$ of the edge scales as $1/{\rm N}$, and the weight
of an interior pixel cannot depend on N, we see that
the weight of an interior pixel must also be independent of $p$.
As a result $\sum_{d_{xy} \in {\cal C}} w_{xy} = \sum_{d_{xy} \in {\cal P}} a^2_{xy}w_{xy}$.

Therefore, we find that

\begin{equation}
{\cal R}^2 = \frac{\sigma_c^2}{\sigma_p^2} = \frac{\sum_{d_{xy} \in {\cal C}} w_{xy}^2 \sigma_{xy}^2}{\sum_{d_{xy} \in {\cal P}} a^2_{xy}w_{xy}^2 \sigma_{xy}^2}
\end{equation}

Although $\cal R$ must be calculated for any given set of dithers, there
is perhaps one case that is particularly illustrative.  
When one has many dithers, and these dithers are fairly
uniformly placed across the pixel, one can  approximate the effect of
the dither pattern on the noise by assuming that the dither pattern
is entirely uniform and continuously fills the output plane.  In this
case the above sums become integrals over the output pixels,  and thus it
is not hard (though somewhat tedious) 
to derive $\cal R$.    
If one defines $r = p/s$ where $p = {\tt pixfrac}$ and $s = {\tt scale}$, 
then  in the case of a filled uniform dither pattern one finds, 

if $r \geq 1$
\begin{equation} {\cal R} = \frac{r}{1-\frac{1}{3r}},
\end{equation}

and if $r \leq 1$,  
\begin{equation} {\cal R} = \frac{1}{1-\frac{r}{3}}.
\end{equation} 

Using the relatively typical values of $p = 0.6$ and
$s = 0.5$,  one finds ${\cal R} = 1.662$.  This formula can also
be used when block summing the output image.  For example, 
a weighted block-sum of $N \times N$ pixels is equivalent to drizzling
into a single pixel of size $Ns$.  Therefore, the correlated noise
in the blocked image can be estimated by replacing $s$ with $Ns$ in 
the above expressions.

\section{SOME EXAMPLES OF THE APPLICATION OF DRIZZLE}

Drizzle has now been widely used for many astronomical
image combination problems. In this section we briefly note some of
these and provide references where further information may be obtained.

Drizzle was developed for use in the original Hubble Deep 
Field North, 
a project to image an otherwise unexceptional region
of the sky to depths far beyond those of previous astronomical images.
Exposures were taken in four filter bands from the near ultraviolet to the
near infra-red.  The resulting images are available in the published
astronomical literature \cite{hdf+96} as well as from the Space Telescope
Science Institute via the World Wide Web at 
http://www.stsci.edu/ftp/science/hdf/hdf.html.  

Subsequently Drizzle has also been applied to the Hubble Deep
Field South \cite{hdf-s-w}. In this case it was used for the combination
of images from NICMOS \cite{hdf-s-nic} and STIS \cite{hdf-s-stisim}  
as well as WFPC2 \cite{hdf-s-wfpc2}. In order to obtain 
dithered NICMOS and WFPC2
images in parallel with STIS spectroscopy,  HST was rotated, as well
as shifted, between observations during the HDF-S.  All the
software developed to handle such challenging observations is
now publicly available (see Section 9).

The HDF imaging campaigns are atypical as they had a large numbers of dither
positions.  A more usual circumstance, matching that described in Section 3,
is the processing of a small number of dithers without multiple
exposures at the same pointing. A good example of such imaging and its
subsequent processing is provided in Fruchter \etal (1999)\nocite{grb-fr-99} 
where Gamma Ray Burst
host galaxies were observed using the
STIS and NICMOS HST cameras to obtain morphological
and photometric information.   Similarly Bally, O'Dell and McCaughrean
(2000) \nocite{bom00} have used Drizzle to combine 
dithered WFPC2 images with single exposures at each
dither position in a program to observe disks,
microjets and wind-blown bubbles in 
the Orion nebula.  

Examination of these published images may help 
the reader to obtain a feeling for the results
of using the Drizzle program. 
In addition an extensive set of worked examples of combining dithered
data using Drizzle is available in the Dither Handbook \cite{drhand}
distributed by STScI.

\section{CONCLUSION}

Drizzle provides a flexible, efficient means of combining dithered
data which preserves photometric and astrometric accuracy, obtains
optimal signal-to-noise, and approaches the best resolution that
can be obtained through linear reconstruction. An extensively
tested and robust implementation is freely
available as an IRAF task as part of the STSDAS package  and
can be retrieved from the Space Telescope Science Institute
web page (http://www.stsci.edu).  In addition to Drizzle, a number
of ancillary tasks for assisting with determining the
shifts between images and the combination of WFPC2 data are
available as part of the ``dither" package in STSDAS.  

We are continuing to improve Drizzle, to increase both ease
of use and generality.    New versions of Drizzle will be incorporated
into STSDAS software updates.    Additional capabilities will
soon make the alignment of images simpler, and will
provide the user with a choice of drizzling kernels, including
ones designed to speed up the image combination with minimal
change to the output image or weight -- an enhancement which 
may prove particularly useful in the processing ACS images.
Although these additions may make Drizzle somewhat more flexible,
the basic algorithm described here will remain largely
unchanged, as it provides a powerful,
general algorithm for the combination of dithered undersampled
images.

\section{ACKNOWLEDGEMENTS}

Drizzle was developed originally to combine the HDF-N datasets.
We wish to thank our colleagues in the HDF-N team, and Bob Williams
in particular, for encouraging us, and for allowing us to be part
of this singularly exciting scientific endeavor.  We also thank
Ivo Busko for his work on the original implementation of the
STSDAS dither package, Hans-Martin Adorf for many entertaining
and thought provoking discussions on the theory of image combination,
and Stefano Casertano for inciting us to develop a more general
theory of the correlated noise in drizzled images.  Finally,
we are grateful to Anton Koekemoer for a careful reading of the
text, and to our referee, Tod Lauer, 
for numerous suggestions which significantly improved
the clarity and presentation of this paper.



\eject

\eject

\end{document}